\documentclass{article}
\usepackage{graphics}
\begin{document}

\newcommand{\bc}{\begin{center}}
\newcommand{\ec}{\end{center}}
\newcommand{\be}{\begin{equation}}
\newcommand{\ee}{\end{equation}}
\newcommand{\no}{\noindent}
\newcommand{\vs}{\vspace{2.5mm}}
\newcommand{\vn}{\vs\no}
\newcommand{\rmd}{{\rm d}}
\newcommand{\rme}{{\rm e}}
\newcommand{\rarr}{\rightarrow}
\newcommand{\x}{\mbox{$\times$}}
\newcommand{\dg}{\mbox{$^\circ$}}
\newcommand{\dn}[1]{\mbox{$_{#1}$}}
\newcommand{\mult}{\multicolumn}

 \baselineskip 6 mm
 \bc

{\LARGE\bf  Rhombic cell analysis - A new way of probing the
large-scale structure of the universe. I. General
considerations}

\vs\vs{\large  T. Kiang}

{\textit{Dunsink Observatory, Dublin Institute for Advanced
Studies,\\ Dublin 15, Ireland}}\ec 

\baselineskip 4.46 mm
\begin{quotation}

\vs\vs\no{\bf Abstract}~~ A new way of probing the
large-scale structure of the universe is proposed. Space is
partitioned into cells the shape of rhombic dodecahedron. The
cells are labelled ``filled'' or ``empty'' according as they
contain galaxies or not. The cell size is so chosen as to
have nearly equal numbers of filled and empty cells for the 
given galaxy sample. Two observables on each cell are
definable: the number of its like  neighbors, n 1,  and a
two-suffixed topological type $\tau$, the suffixes being the
numbers of its  like and unlike neighbor-groups. The
frequency distributions of  $n_1$  and $\tau$ in the observed
set of filled (empty) cells are then considered as indicators
of the morphology of the set. The method is applied to the
CfA catalogue of galaxies as an illustration. Despite its
limited size, the data offers evidence 1) that the empty
cells are more strongly clustered than the filled cells, and
2) that the filled cells, but not the empty cells, have a
tendency to occur in sheets. Further directions of
development both in theory and application are
indicated.       
   
\vs\vs\no{\bf Key words:}~~cosmology: large-scale
structure---cosmology: observations---CfA Catalogue

\end{quotation}

\bc{\bf 1. INTRODUCTION}\ec

   This paper, hopefully, will open up a new line of research into 
the large scale morphology of the universe by means of rhombic 
dodecahedron cells. In this first paper, I shall outline some of the 
general principles, and, by way of illustration, apply the analysis 
to the CfA Catalogue of Galaxies (Davis and Huchra 1982), to address 
the question of the shape of the filled/overdense and 
empty/underdense regions of the universe.  Further development, both 
in theory and application, are being actively pursued, and hopefully 
will appear in print in the near future. A much shorter, first draft 
of the present paper was published some 10 years ago (Kiang 1993). 
The journal in which it appeared, however, had a rather limited 
circulation, particularly so for researchers in China. 

\newpage
\bc{\bf 2. THE LARGE-SCALE MORPHOLOGY}\ec

    The classic book on this subject is Hubble's ``The Distribution 
of the Nebulae'' (Hubble 1936). In it the author articulated the 
thesis that the galaxies are the building blocks of the universe and 
that, at least as a first approximation or a working hypothesis, we 
should take them as being statistically uniformly distributed in 
space. But the working hypothesis of an authority has a tendency of 
persisting as a literal truth in the minds of the later researchers, 
and it was not until the 1950s that Hubble's idea of uniform 
distribution was seriously challenged. The challenge came from two 
directions: on one hand, Neyman and Scott (1952, 1953b), using 
statistical techniques based on counts-in-cells, showed that the 
galaxy distribution is statistically clustered  rather than 
statistically uniform, and on the other, Abell actually identified 
from Palomar Sky Survey plates, 2712 operationally well-defined rich 
clusters of galaxies (Abell 1958). (Of course, Abell's catalogue and 
its Southern extension have since acquired the status of a classic 
database). The next step in the direction further away from the 
uniform distribution was taken by myself (Kiang 1965, 1971): I 
applied the counts-in-cells technique to the Abell clusters and found 
they themselves were clustered. This led me to the idea that galaxies 
may be clustered on all scales. About the same time, Peebles began a 
different way of characterising the galaxy distribution, a way which 
proved to be highly productive, namely, the calculation of the 
two-point correlation function (Peebles 1978). The two-point correlation 
function is in some sense a quantitative expression of my idea of 
indefinite clustering.  As is well-known, the calculation of the two-
point correlation, and less often, of the three-point correlation 
function has become something of an industry in the field of galaxy 
research. Among the large number of applications, two of the more 
recent examples may be cited (Zhu 1997a, 1997b).   

    All the studies mentioned so far share the common belief that the 
galaxies are the basic constituents of the universe, our exclusive 
concern when dealing with the question of the large-scale morphology.  
This outlook seemed so natural. However, a glimpse of an alternative 
did appear in a paper presented at the 1977 IAU Symposium in Tallinn, 
tellingly entitled ``Has the Universe the Cell Structure ?''.  And it 
may not be coincidental that, at the same symposium, Zel'dovich gave 
a paper on his well-known pancake theory of galaxy formation 
(Zel'dovich 1978), for Zel'dovich's theory would go hand in glove 
with a cell structure. Since then, large voids or underdense regions 
and structures like the ``Great Wall'' have become common knowledge. 
Thus, the idea dawned that voids may be as a fundamental ingredient 
of the large-scale structure as are the galaxies. 

   The first paper that treated the underdense and overdense regions 
on an equal footing was by Gott et al. (1986). These authors found (i) 
that the two regions are each a sponge-like, connected entity, and 
(ii) that they are equivalent. I think that statement (i) is entirely 
plausible, but to give statement (ii) any quantitative interpretation 
should perhaps await a more penetrating analysis than what they did: 
they partitioned space into cubic cells and examined the interface 
between the two regions. It is my belief that partitioning space into 
what I call rhombic cells would provide a much more powerful means of 
analysis, and that the analysis should not be confined to the 
interface, but should extend to all the constituent cells.  I hope 
the present paper will offer a first glimpse of the great richness 
that is inherent in the rhombic cell analysis.

\bc{\bf 3. RHOMBIC CELL ANALYSIS}\ec

   I imagine space is partitioned into cells the shape of rhombic 
dodecahedron. The latter can be imagined to form in the following way: 
1. Space is partitioned into identical cubes, and the cubes are 
painted alternately black and white into a three-dimensional 
chessboard. 2. Each white cube is cut into six identical pyramids 
with the faces as the bases and the centre as the common vertex. 3. 
To a given black cube, we stick on the six adjoining white pyramids. 
The result: a solid with 12 identical white rhombic faces generated 
out of a black cube. I shall call the black cube the generating cube 
of the dodecahedron. The size (volume) of the dodecahedron is, of 
course, twice the size of its generating cube. The identical rhombic 
face has its semi-minor diameter, semi-major diameter and side in the 
ratios of 1:$\sqrt{2}$:$\sqrt{3}$. More of the geometry of the rhombic 
dodecahedron is given in the  Appendix at the end of this paper. The 
minor diameters of the 12 faces are just the 12 edges of the 
generating cube;---this one-to-one correspondence was made use of in 
my construction of the Table 1 below. 

   Thus, space is partitioned into such 12-rhombic-faced cells. From 
now on, these cells will simply be referred to as ``rhombic cells''.  
An analysis of the CfA Catalog of Galaxies (Davis and Huchra 1982) 
now follows, both for its own sake and as an illustration of method: 
all the definitions and arguments will equally apply to any other 
galaxy sample. If a rhombic cell contains no CfA galaxies, then we 
call it an empty cell; otherwise, it is a filled cell. The ensemble 
of the empty cells forms the empty region, and similarly the filled 
region. So the question we are concerned with here is ``Do the filled 
and empty regions have the same morphology ?'', or more precisely, 
``granted that they are each a connected entity, do they have the 
same mix of 1-dimensional string-like, 2-dimensional sheet-like, and 
3-dimensional clump-like contents ?''

\vn{\bf 3.1 Optimal Cell Size} 

   Obviously, the question is at its most meaningful when the
two regions have the same size, for the morphology of a
region clearly depends on its overall size. The size (volume)
of the cell when this happens will be called the optimal cell
size, $v_{\rm optm}$. It is easily shown that $v_{\rm optm}$ 
always exists for any reasonably spread-out distribution of
the galaxies. Suppose we have a sample of $N$ galaxies 
distributed throughout some volume of space $V$. We enclose
each galaxy in a small sphere, the result: $N$ tiny spheres
in a huge empty sea: the two ensembles differ completely in
morphology and size. Now let us keep on increasing the size
of the spheres. The spheres will begin and keep on
coalescing, but independently of any merging, provided the
galaxies are reasonably spread out to start with, the total 
volume of the filled region, $V_{\rm filled  region}$, will
keep on growing at the expense of the total volume of the
empty region, $V_{\rm empty region}$, and we just stop the
process when the two become equal. This demonstrates the
``mathematical'' existence of $v_{\rm optm}$. Its practical
evaluation can always done by trial and error, but a good 
starting value could be $v_{\rm optm} = \ln 2 . (V / N )$ ---
the value that would result from a Poisson distribution of
the number of galaxies per cell.

\vn{\bf 3.2 Test Statistics} 

   Here we come to the heart of the present analysis. The
issue is, given an ensemble of filled or empty cells, what
observables or functions of observables of the individual
cells can we identify whose number distributions in the given
ensemble can serve as probes of the morphology of that
ensemble ?  

\vn{\it 3.2.1 The Number of Like Neighbors $n_1$} 

   If two neighbor cells are both filled or both empty, then
their common face will be called an {\em inner wall\/}; otherwise, an
{\em outer wall\/}. Let, for a given cell, $n_1$  be the number of
its inner walls ($n_1  =  0, 1, 2, \cdots, 12$). Now consider
the frequency distribution or number distribution of $n_1$
for the whole filled region, $N_{\oplus}(n_1)$. Since we know
the galaxies are not distributed in space purely at random, we
can expect $N_{\oplus}(n_1)$ to depart from the well-known 
binomial form. And the same is true for the similarly defined
$N_{\bigcirc}(n_1)$ for the empty region.  Each of the two
distributions of $n_1$, then, reflects some non-random
features of its own region, and it may be reasonable to 
expect that any difference between their morphologies may in
turn be reflected in some difference between the two
distributions. However, it should be noted here that the
difference between the mean $n_1$ - values of the two
regions, $<n_1>_{\oplus}-<n_1>_{\bigcirc}$ tells us nothing
about the intrinsic morphologies, because any non-zero
difference in the mean value will be entirely due to
accidental contributions from the ``boundary'' walls and to a
departure from exactly equality between the numbers of filled
and empty cells. (This statement can be easily proved
starting from the fact that an outer wall for a filled cell is
also an outer wall for an empty cell, and the numbers of outer
and inner walls of any cell must add up to 12). Thus, only
differences between $N_{\oplus}(n_1)$ and
$N_{\bigcirc}(n_1)$  in respect of some parameters other than
the mean value are useful to us.

\vn{\it 3.2.2  The Topological Type $\tau_{m_1, m_2}$} 

      Consider a cell with $n_1$ inner walls and
$n_2(=12-n_1)$ outer walls. Adjacent inner walls are said to
form an {\em inner wall group\/}; similarly, an 
{\em outer wall group\/}. A
cell with $m_1$ inner wall groups and $m_2$ outer wall groups
is then said to be of topological type $\tau_{m_1,m_2}$. 

     A single cell surrounded by 12 unlike neighbors is of
type $\tau_{0,1}$. For further correlations, it is convenient
to adopt the following nomenclature. An agglomeration of like
cells which is at least 3-cells thick in at least one
dimension will be called a thick clump. A string of single
cells is called a single strand, a sheet one cell thick, a
mono-layer. A 2-cell-thick string is a 2-ply, and a 
2-cell-thick sheet, a double-layer. The following statements
are then obvious:  only the inside cells of thick clumps are
of type $\tau_{1,0}$; only the cells of single strands are of
type $\tau_{2,1}$; only the cells of mono-layers are of type
$\tau_{1,2}$; while the boundary cells of thick clumps, and
the cells of double-layers and of 2-plys are all $\tau_{1,1}$.
Thus, the number distribution of $\tau$-types can only give a
rough indication of the mix of various types of objects.
However, when the data is of limited size, the univariate
$\tau$-distribution may be all that we can work with.

\vn{\it 3.2.3.  The Reference Bivariate $(n_1,\tau)$
Distribution} 

     As already mentioned, in the case of pure random
distribution of filled and empty cells (ie, when a given cell
has an equal probability of on-half of being filled or
empty), the univariate $n_1$- distribution is the binomial
distribution with parameter one-half. The univariate
$\tau$-distribution in the random case is, however, unknown 
and must be evaluated {\it ab initio\/}.  While attempting to
do this, I realised that I could as easily evaluate the
bivariate $(n_1,\tau)$- distribution at the same time. And
the latter may in any case be required when larger samples
become available.  Distributions for the random case will be
used as reference and will be labelled as such; this
particular one will be denoted by $N_{\rm ref} (n_1,\tau)$.

    My evaluation of $N_{\rm ref} (n_1,\tau)$ was laborious.
First, recalling the (1,1) - correspondence between the cell
faces and the edges of its generating cube, any mix of inner
and outer walls of a cell is equivalent to the same mix of
two types of edges of a cube (type-1 for the inner, type-2
for the outer, say). Now, we must distinguish between a
complexion and a configuration. A complexion is any 
assignment of either 1 or 0 to each of the 12 edges (regarded
as distinct or labelled) of the cube. There are altogether 
$4096 (=2^{12})$ complexions. A configuration, on the other
hand, is any relative arrangement of the two types of edges
on the cube. A configuration generally corresponds to $f$ 
complexions, with $f$  taking the values, in order of
increasing frequency, 4, 6, 12, 24 or 48 (the last if mirror
images are considered as one configuration). Now, each 
configuration has a unique value of $n_1$, and a unique
$\tau$-type. So the given configuration adds $f$  to the
frequency in the $(n_1,\tau)$-box. Going through all the
configurations results in the entire distribution $N_{\rm
ref} (n_1,\tau)$. And that is all there is to it;---in 
principle. In practice, it is often difficult to be sure (1)
that indeed all configurations have been examined, and (2)
that the configurations examined contained no duplications.
Another source of error lurks in the evaluation of $f$: all
symmetries in the configuration must be noted and allowed
for, otherwise $f$  will be grossly overestimated. I found
the Schlegel diagram representation of the cube (a plane
graph which preserves only the topology of the cube) 
convenient for determining the $\tau$-type of a given
configuration, and the ``unfolded cube'' (where the faces are
opened out along common edges onto the same plane) sometimes
useful for checking whether a given configuration has not
already been examined and for recognizing any symmetry it
might possess. Having now detailed the difficulties, I now
mention a welcome simplification: we need only evaluate for 
values of $n_1$ between 0 and 6. This can be shown as follows.
Consider a $(n_1,\tau_{m_1, m_2})$-configuration. By
interchanging the inner and outer walls (or the type-1 and
type-2 edges) we obtain a $(12-n_1, \tau_{m_2,
m_1})$-configuration. The two configurations have one and the
same $f$-value, so this pair of ``mirror configurations''
make the same contribution $f$  to their respective $(n_1
,\tau)$-boxes. But all the configurations can be so paired,
hence we have, generally, $ N_{\rm ref}(n_1,\tau_{m_1,m_2}) =
N_{\rm ref}(12-n_1,\tau_{m_2,m_1})$. The result of my 
evaluation some 10 years ago is given in Table 1. I have
recently repeated the evaluation and confirmed the results.

      I suggested to Mr Y. F. Wu (Wu Yongfeng) of the Center of  
Astrophysics, University of Science and Technology of China, Hefei, 
that he might make an independent evaluation on the computer. He 
wrote a program using the C language and duly verified the entire 
table. We can now be confident that all the numbers given in Table 1 
are correct.

\begin{center}
{\bf Table 1  The Reference Bivariate Distribution 
$N_{\rm ref}(n_1,\tau)$}

\vs
\small
\begin{tabular}{c|rrrrrrrrrrrrr|r}

$\tau$  & \mult{13}{c}{$n_1$} & N\dn{ref}[$\tau$] \\ \cline{2-14}

\{$m_1, m_2$\} & 0 & 1 & 2 & 3 & 4 & 5 & 6 & 7 & 8 & 9 & 10 &
11 & 12 & \\ \hline

 \{1,1\} &   & 12 & 24 & 56 & 126 & 252 & 340 & 252 & 126 &
 56 & 24 & 12 & & 1280 \\ 

 \{2,1\} & & & 42 & 120 &252 & 336 & 160 & 24 & & & & & & 934
\\

 \{1,2\} & & & & & & 24 & 160 & 336 & 252 & 120 & 42 & & &
934 \\

 \{2,2\} & & & & & 12 & 72 & 240 & 72 & 12 & & & & & 408 \\

 \{3,1\} & & & & 44 & 96 & 72 &&&&&&&& 212 \\

 \{1,3\} &&&&&&&& 72 & 96 & 44 &&&& 212 \\

 \{3,2\} &&&&&&	36 & 12 &&&&&&& 48 \\

 \{2,3\} &&&&&&& 12 & 36 &&&&&& 48 \\

 \{4,1\} &&&&& 6 &&&&&&&&& 6 \\

 \{1,4\} &&&&&&&&& 6 &&&&& 6 \\

 \{4,2\} &&&&& 3 &&&&&&&&& 3 \\

 \{2,4\} &&&&&&&&& 3 &&&&& 3 \\

 \{0,1\} & 1 &&&&&&&&&&&&& 1 \\

 \{1,0\} &&&&&&&&&&&&& 1 & 1 \\ \hline

 N\dn{ref}[$n_1$] = & 1 & 12 & 66 & 220 & 495 & 792 & 924 &
792 & 495 & 220 & 66 & 12 & 1 & 4096 \\  
 \end{tabular}
 \normalsize
 \end{center}

\vn{\bf 3.3  The Distance Effects}

     There are two distance effects due to properties inherent in 
astronomical surveys. First, since every astronomical survey covers 
some region on the sky, we make the most efficient use of the data 
when we line up our rhombic cells in fixed solid angles. This means 
that, in order that all the cells have the same size (volume), their 
radial dimension must be inversely proportional to the square of the 
distance. If we set the generating cube of the cell at some far, 
fiducial distance to be more or less a cube, ie, with all its sides 
more or less equal, then as we move closer, the generating cube will 
depart further and further from being a cube, its two transverse 
dimensions each decreasing as the distance $r$, and its radial 
dimension increasing as $r^{-2}$, and the rhombic cell distorts 
correspondingly. 

     The second effect is due to the fact that, in respect of
statistical completeness, we can at best have surveys complete
to some apparent magnitude $m$.  For the CfA Catalogue, $m =
14.5$. Then suppose we consider only galaxies brighter than
absolute magnitude - 15.5 (which comprises 96\% of all the
galaxies in the CfA catalogue), then the catalogue is
complete only up to distance $r = 10$ Mpc and is increasingly
incomplete after that. The completeness factor at distance
$r$, ($r > 10$ Mpc), is

\be
s(r) = \frac{\int^{M(r)} \Phi(M)\rmd M}
            {\int^{-15.5} \Phi(M)\rmd M} \;,
\ee
\no where

\be
M(r) = 14.5 - 5 \log (r/{\rm Mpc}) - 25 \;,                                     
\ee 
\no and $\Phi(M)$ is the adopted luminosity function of
galaxies. Now, the number of CfA galaxies lying within 10 Mpc
is very small: to obtain results of any statistical
significance we must include those lying beyond, but in doing
so, we must increase the cell size by $s(r)^{-1}$ so as to
maintain the same mathematical expectation of the number of 
galaxies per cell. The two distance effects together mean that
the radial separation between the cell centres should vary as
$1/(r^2.s)$. And there is some advantage in taking 10 Mpc as
the fiducial distance, thus:  

\be
\Delta r = \Delta r(r) = 
\Delta r_{\rm 10 Mpc} / [(r / {\rm 10 Mpc})^2 . s (r)] \;,
\ee
\no where $\Delta r_{\rm 10 Mpc}$ is the value of $\Delta r$
at 10 Mpc, and is set equal to $v_{\rm optm}/2)^{1/3}$ and
$s(r)= 1$ for $r < 10$ Mpc, and is given by (1) for $r > 10$
Mpc. The advantage is that this way we get the greatest
number of the least distorted cells.

      In practice, of course, we may cut off our sample at some 
reasonable value of $s$, 0.2 or 0.1, say.

\vn{\bf 3.4  The Boundary Effect}

      Because of the shape of the rhombic cell, we have cells with 
the property that, while their centres lie inside the region covered 
by the data, parts of them lie outside. Such cells will be called 
``boundary cells''. If a boundary cell is a filled cell, then we can 
be sure that it is a filled cell, but if it is an empty cell, then we 
cannot be sure of its empty status without knowing whether there are 
any galaxies in its outlying part. In either case we do not know the 
inner-wall/outer-wall status of any ``boundary faces'' (faces that 
adjoin outside cells). It might be suggested that we should simply 
discard all empty boundary cells, but this is not sufficient, for the 
inner/outer status of the walls of any cell next to any one of them 
is still uncertain. So we discard all cells next to any empty 
boundary cells as well. The result would be a very jagged working 
region. This is generally undesirable, so we discard all the boundary 
cells, and all the cells in the ``next shell'': the consequent 
wastage might then prove unacceptable. 

      There is a simple way out. Imagine the entire observed sample 
of galaxies is bodily repeated along the two transverse coordinate 
directions. Then, all the boundary cells will have a sure 
empty/filled status, and all the boundary walls, a sure inner/outer 
status. Of course, some spurious correlation will be introduced, but 
unless the region surveyed is exceptionally narrow,the effect of any 
such correlation can be expected to be unimportant.

\bc{\bf 4.  APPLICATION TO THE CfA CATALOGUE}\ec

\begin{figure}
\resizebox{\columnwidth}{!}{\includegraphics*{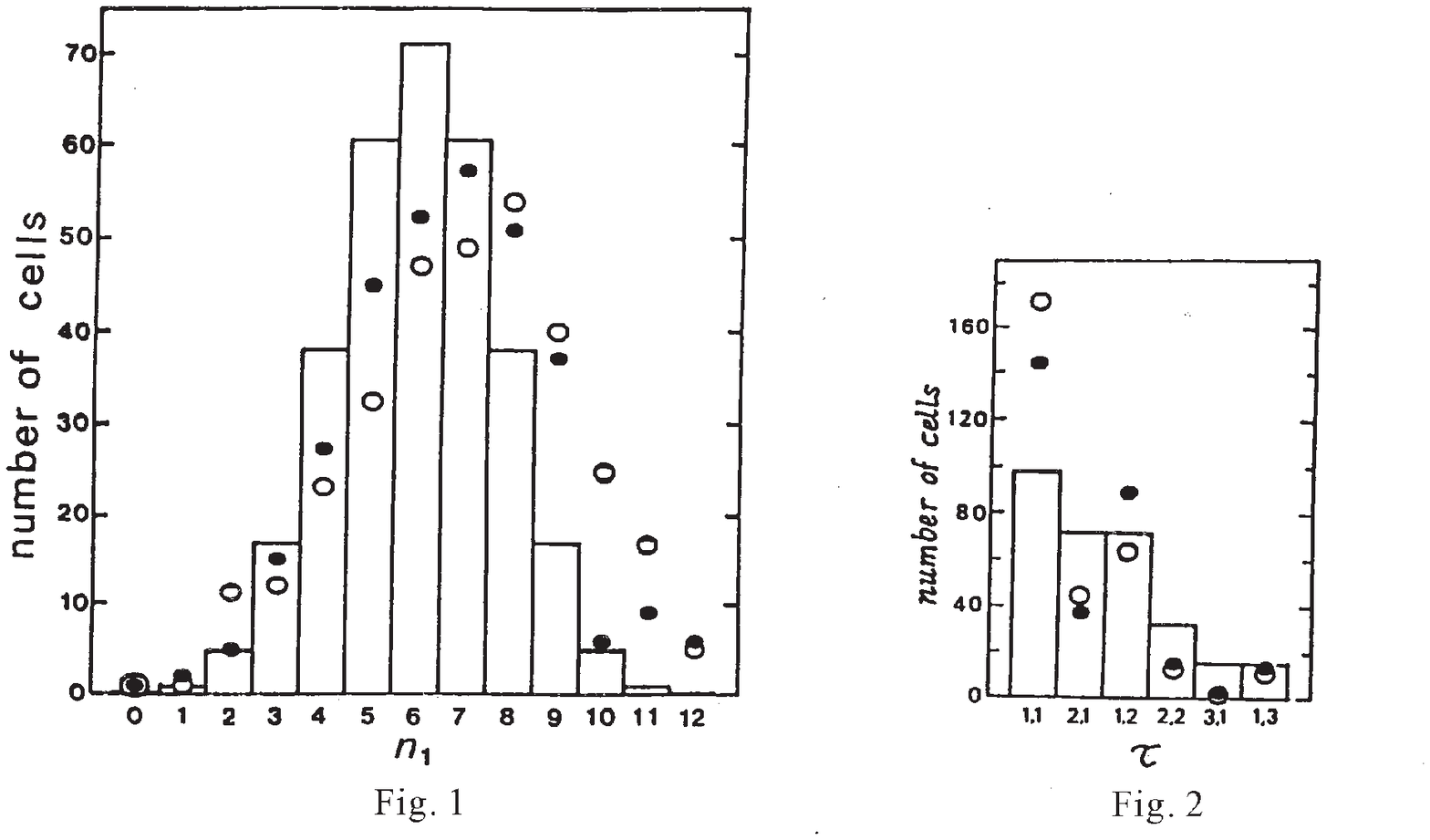}}
\bc
{\small Left: Fig.\,1. Frequency distributions of the number
of like neighbors $n_1$ for filled cells (filled ellipses)
and empty cells (empty ellipses). The histogram is the
reference (binomial) $n_1$-distributon of Table 1, normalised
to the mean observed total, 315.\\
Right: Fig.\,2. Mariginal distributions of the topological
type $\tau$ (part only). Symbols have the same meaning as in
Fig.\,1 and the histogram is the reference $\tau$-distribution
of Table 1, normalised to a total of 315.} 
\ec
\end{figure}

      For the luminosity function in equation (1) I followed
Davis and Huchra (1982) and took it to be the Schechter
function with $\alpha = - 1.3$ and $M_{\ast}=-19.4$. Then
after a few trials I found that if we set  $r_{\rm 10 Mpc} =
2.2$ Mpc (the Hubble constant is set to be 100 (km/s)/Mpc 
throughout this paper), then the northern sample of the CfA
will give 262 empty and 248 filled cells, and the southern
sample, 55 empty and 65 filled cells. The combined sample
would then give 317 empty and 313 filled cells;---probably as
equal a pair as we could ever get.

      Incidentally, the ratio in solid angle between the two samples 
is $0.83/1.83 =0.45$ (Davis and Huchra 1982), but the ratio in the 
number of usable cells is only $(55+65)/(262+248) = 0.24$. This is 
because the southern sample occupies an elongated strip in the sky 
and its poor return is in spite of the fact that I had moved the 
origin of its equal-area projection to somewhere near the middle of 
the strip. 

      Because of the limited size of the sample, results will be 
given only for the two marginal distributions. Fig.\ 1 displays the 
observed $n_1$-distributions (the filled and empty ellipses) against 
the reference binomial distribution (the histogram), and Fig.2 
displays similarly the observed and reference $\tau$-distributions. 

Fig.\ 1 shows that the two observed distributions depart
greatly from the binomial form. This is not surprising since
we know the galaxy distribution in space is clustered rather
than statistically uniform. But what is perhaps surprising is
that the degree of clustering seems to be higher in the empty
region than in the filled region. This statement is based on
the following features not involving the mean value (cf. the
caveat in 3.2.1): 1) The modal $n_1$ is 8 for the empty
region as against 7 for the filled region. 2) The fraction of
cells with $n_1 = 10$ for the empty region is 8\%, higher 
than the 5\% for the filled region, while the fraction of
cells with $n_1 = 5$ for the empty region is 10\%, lower than
the 14\% for the filled region. 

      Fig.\ 2 shows the following notable features. 1) The
observed numbers of $\tau_{1,1}$-type cells in both the empty
and filled regions greatly exceed the random expectation,
while the observed numbers of $\tau_{2,2}$-, $\tau_{3,1}$-,
$\tau_{1,3}$-type cells in both regions fall short of the
random expectations. Both these features are consistent with a
tendency for likes to stick together for both filled and empty
cells.  2) That the observed number of $\tau_{1,1}$-type
cells is higher in the empty than the filled region is
probable evidence for a greater clustering tendency in the
former. However, in view of the analysis of the
$\tau_{1,1}$-type cell in 3.2.2, a firmer and more precise 
conclusion must await an analysis of the array or conditional
distribution $N (n_1 | \tau_{1,1})$. 3) Both regions tend to
avoid single strands ($\tau_{2,1}$-type cells). 4) The filled
region, but not the empty region, seems to have more
$\tau_{1,2}$-type cells than random expectation: the
galaxies, but not the voids, seem to have a tendency of
occurring in very thin sheets (mono-layers). This is a rather
remarkable result, and should certainly be checked further
with larger size data.

\bc{\bf 5. PROSPECTS OF FURTHER DEVELOPMENT}\ec

     There is great scope for development both in the theory of 
rhombic cell analysis and in its applications.

     Much larger survey results than the CfA Catalogue are now 
available. To each of these, we can, of course, apply the present 
method of analysis. But more than that, we can refine the analysis in 
two ways:

     1. Recall that, because of the small size of the CfA Catalogue, 
I had to include the more distant regions where the catalogue is 
incomplete, by the device of increasing proportionately the size of 
the cells (Section 3.1). But this procedure implicitly assumes that 
the morphology in the near space on a certain standard scale (the 
optimal cell size) is the same as the morphology in the far space at 
larger and varying scales. When much larger datasets become available, 
we might be able to investigate the near and far regions separately, 
thus instead of making the assumption that the morphology is scale 
invariant, we actually investigate it.

     2. Again, because of the limited sample size, this paper
only examined the two marginal distributions of $n_1$ and 
$\tau_{m_1,m_2}$. When the sample is sufficiently large, we
can examine the conditional distributions of $n_1$  at a
given $\tau$.  The conditional distribution of $n_1$  at
$\tau_{1,1}$ is of particular interest: it is by far the
largest of all such conditional distributions, hence most
amenable  to a finer analysis, and by its analysis we may be
able to estimate what fraction of the large number of 
$\tau_{1,1}$-type cells come from what source, thick clumps,
double-layers or 2-plys. It would be most interesting if the
double-layer turns out to be favorite form for the filled
region, since we already seem to have evidence that the mono-
layer is so (see end of Section 4).      

      As regards theoretical development, there is a
completely new dimension to be explored. The present analysis
is of cells considered as individuals. Now, for each cell
there is a natural definition of its (like) neighbors of
ranks 1, 2, 3, etc. And there should be many powerful test
statistics based on properties of entire neighborhoods of
various ranks. But even within the field of individual cells,
we can (i) raise the threshold for the filled cell to 2 or
more galaxies, so that the resulting filled and empty regions
will be closer to being underdense and overdense regions, and
(ii) search for new test statistics besides $n_1$ and
$\tau_{m_1,m_2}$.  Of course, any one of the theoretical
developments could be applied to any one of the new datasets,
and conversely each new application could suggest new
additions to the store of test statistics. The scope for
development is very great indeed.

\vn{\bf Acknowledgement}  I thank Dr X.-Y. Xia  (Xia Xiao-yang) for
providing me with an electronic copy of the CfA Catalogue in
the early days of this work. I also thank Mr. Y. F. Wu (Wu
Yongfeng) for checking the contents of Table 1 with a computer
program.

\newpage
\bc {\bf APPENDIX} \\

{\bf Geometry of the Rhombic dodecahedron}\ec

\begin{figure}
\resizebox{\columnwidth}{!}{\includegraphics*{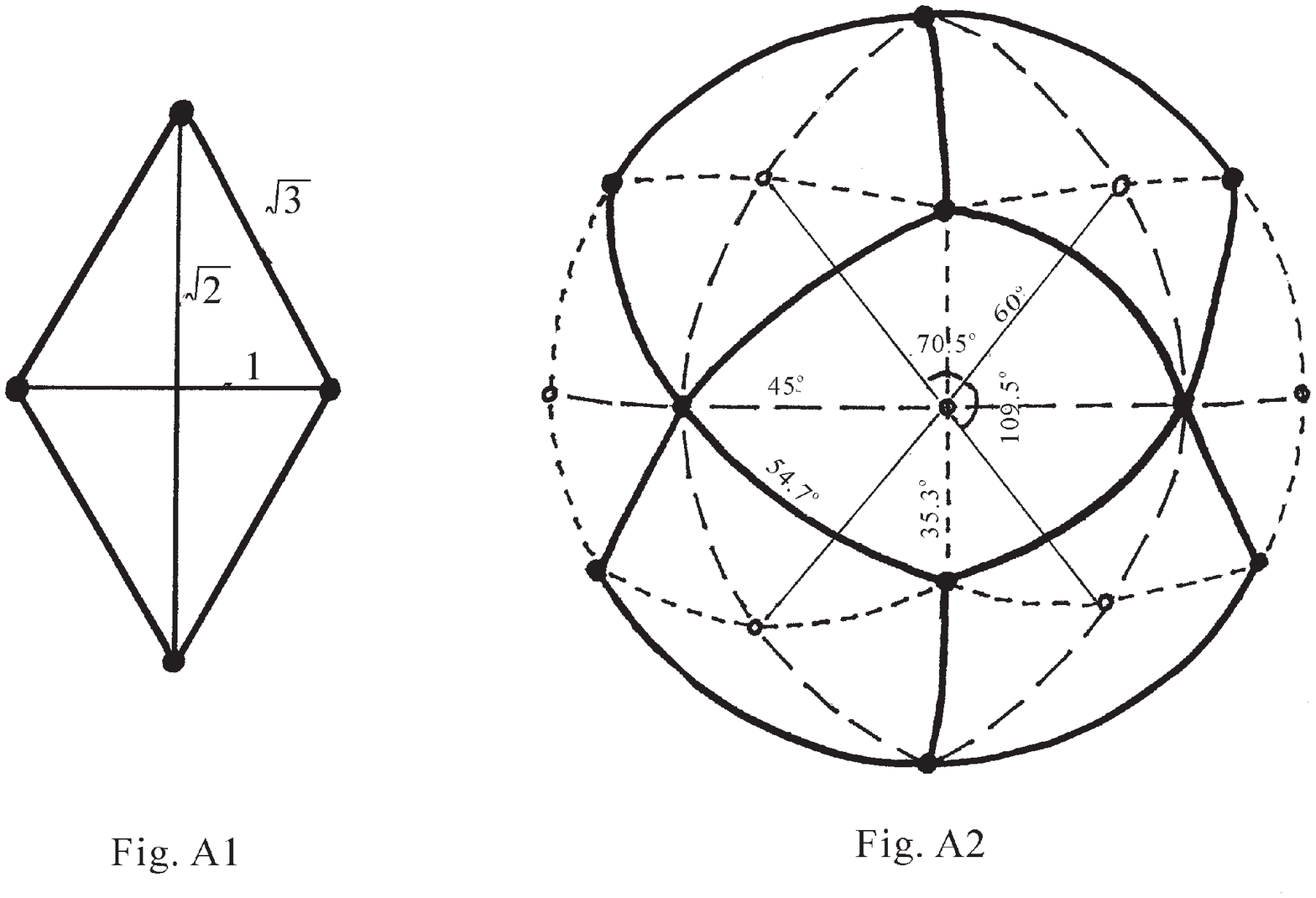}}
\bc
{\small Left: Fig.\,A1; right: Fig.\,A2}
\ec
\end{figure}

      This Appendix lists some of the regular and semi-regular
properties of the rhombic dodecahedron. The dodecahedron has
12 identical rhombic faces. Fig.\,A1 shows the linear
dimensions of the rhombic face, in units of its semi-minor
diameter. The 12 minor diameters of the dodecahedron coincide
with the 12 edges of its generating cube. The dodecahedron
has six 4-vertices (where 4 faces meet) corresponding to the
six face-centres of the generating cube, and eight 3-vertices
(the eight corners of the generating cube). Fig.\,A2 is the
apparent view of the dodecahedron seen from its centre. The
12 faces each subtends a solid angle of $\pi/3$, each has four
sides of 54.7\dg, a major diameter of 90\dg\ and a minor
diameter of 70.5\dg. The arrangement of the faces can be
described as follows. Choose any pair of opposite 4-vertices
as north and south poles, then four faces meet at the north
pole, four at the south pole and four lie length-wise along
the equator. A given face has four neighbors. With respect to
the centre of a given face, the centres of its four neighbors
are all 60\dg\ away, but they are not isotropically placed: the
directions to two neighbors sharing a 3-vertex with the given
face include an angle of  70.5\dg, while the directions to
two neighbors sharing a 4-vertex, one of 109.5\dg.

\end{document}